\documentclass{ws-jmrr}
\usepackage[sort,compress,super]{cite}
\usepackage[caption=false]{subfig}
\usepackage{floatrow}
\usepackage{tabularx}
\usepackage{color,soul}
\usepackage{graphicx}
\usepackage{mathtools}
\usepackage{siunitx}
\usepackage{array}
\usepackage{url}
\usepackage{hyperref}
\usepackage{lineno}
\nolinenumbers

\usepackage{xcolor,cite,etoolbox}
\makeatletter 
\pretocmd\@bibitem{\color{black}\csname keycolor#1\endcsname}{}{\fail}
\newcommand\citecolor[1]{\@namedef{keycolor#1}{\color{blue}}}
\makeatother

\usepackage{comment}
\usepackage[normalem]{ulem}

\includecomment{ET}

\newcommand{\movefrom}[1]{}
\newcommand{\del}[1]{}
\newcommand{\add}[1]{#1}

\begin{ET}
    \renewcommand{\del}[1]{\textcolor{red}{\sout{#1}}}
    \renewcommand{\add}{\textcolor{blue}}
\end{ET}

\begin{document}


\catchline{0}{0}{2019}{}{}

\markboth{Hemanth Manjunatha,
        Shrey Pareek,
        Amirhossein H. Memar,
        Thenkurussi Kesavadas,
        Ehsan T. Esfahani}{Effect of Haptic Assistance Strategy on Mental Engagement in Learning Fine Motor Tasks}

\title{Effect of Haptic Assistance Strategy on Mental Engagement in Fine Motor Tasks}

\author{Hemanth Manjunatha$^a$, Shrey Pareek$^b$, Amirhossein H. Memar$^a$, Thenkurussi Kesavadas$^b$, Ehsan T. Esfahani$^a$\footnote{Corresponding author}}

\address{$^a$Department of Mechanical and Aerospace Engineering\\ 
University at Buffalo, Buffalo, NY, 14260 USA\\
E-mail: hemanthm@buffalo.edu; ahajiagh@buffalo.edu; ehsanesf@buffalo.edu}

\address{$^b$Department of Industrial and Enterprise Systems Engineering\\
The University of Illinois at Urbana-Champaign, Champaign, IL, 61820 USA\\
E-mail: spareek2@illinois.edu, kesh@illinois.edu}
\maketitle

\begin{history}
\received{(October 1st, 2019)}
\revised{(February 12th, 2020)}
\revised{(April 17th, 2020)}
\accepted{(May 1st 2020)}
\end{history}

\begin{abstract}
This study investigates the effect of haptic control strategies on a subject's mental engagement during a fine motor handwriting rehabilitation task. The considered control strategies include an error-reduction (ER) and an error-augmentation (EA), which are tested on both dominant and non-dominant hand. A non-invasive brain-computer interface is used to monitor the electroencephalogram (EEG) activities of the subjects and evaluate the subject's mental engagement using the power of multiple frequency bands (theta, alpha, and beta). Statistical analysis of the effect of the control strategy on mental engagement revealed that the choice of the haptic control strategy has a significant effect (p $<$ 0.001) on mental engagement depending on the type of hand (dominant or non-dominant). Among the evaluated strategies, EA is shown to be more mentally engaging when compared with the ER under the non-dominant hand.
\end{abstract}


\section{INTRODUCTION}
Haptic-based rehabilitation systems are one of the main alternatives to traditional rehabilitation for home-based settings. Sensing the user’s movements, the haptic systems can provide force feedback or guide subsequent motions in a predefined trajectory \cite{oblak2010universal, huq2012development}. In home-based rehabilitation, they can mimic a therapist's actions by providing assistance as needed \cite{passenberg2011towards, pareek2018development, parasuraman2011robot} during fine-motor learning. There are various studies in the literature that report the efficacy of the haptic system for rehabilitation. For instance, Turolla et al. \cite{turolla2013haptic} conducted a haptic-based rehabilitation study involving 15 stroke patients. They reported that all patients showed significant improvements in the clinical (Fugl-Meyer upper extremity scale) and kinematic outcomes (i.e., task completion time, velocity along the traced path, and jerk). To make a comparison between the efficacy of the conventional and robotic rehabilitation, Veerbeek et al. \cite{veerbeek2017effects} conducted a scoping review and concluded that in general, robot-assisted therapy outcomes were better in terms of motor recovery, activities of daily life, strength, and motor control. This was mostly because of the higher training intensity in robot-assisted therapy. However, for the same level of therapy intensity, there was no significant difference between conventional and robotic rehabilitation outcomes. It is very well established that the outcome of the robotic rehabilitation depends on two main factors: the intensity of the exercise and the mental engagement (active participation) of the patient in the therapy \cite{blank2014current, simonetti2016multimodal}.

The training strategies in haptic-based rehabilitation mainly fall into two categories: 
1) \textit{error-reduction (ER)}, which decreases the performance error by providing active assistance to enable the patient to perform the rehabilitation tasks better. 2) \textit{error-augmentation (EA)} that increases the task difficulty to evoke a higher voluntary involvement of the patient to accomplish the goal. 

Despite the wide implementation of both ER and EA strategies, there is still a lack of agreement on which strategy evokes more clinically-significant outcomes after training \cite{li2018effects, israely2016error}. In this regard, Youlin et al. \cite{li2018effects} performed a comprehensive review of the effect of ER and EA strategies in enhancing upper extremity performance and post-stroke recovery. Their review suggested that the EA strategy was statistically more effective than conventional repetitive practice in both motor recovery and task performance. They even reported a statistically significant improvement in motor performance using EA when compared with ER. However, neither EA nor ER evoked clinically significant changes in motor recovery and function \cite{li2018effects, kao2013effect, israely2016error}. 

Regardless of the type of haptic training strategy (EA or ER), the outcome of the haptic-based therapy may not necessarily be superior to manual therapy unless there is an active engagement from the patient \cite{blank2014current, takeuchi2013rehabilitation}. For instance, in the ER, a high level of assistance may render the task too easy, causing the loss of patient engagement and failure to learn the motor primitives\cite{Secoli2011}. Conversely, in EA, the task can become very difficult to the point that it induces anxiety, causing the patient to give up rehabilitation training at an early stage. Therefore, quantification of patient engagement has a pivotal role in the success of robotic rehabilitation. This quantification is even more critical in home-based therapy as the clinician may not be physically present to encourage the patients to perform the rehabilitation tasks.

\begin{figure}[!bht]
    \includegraphics[width=0.85\linewidth]{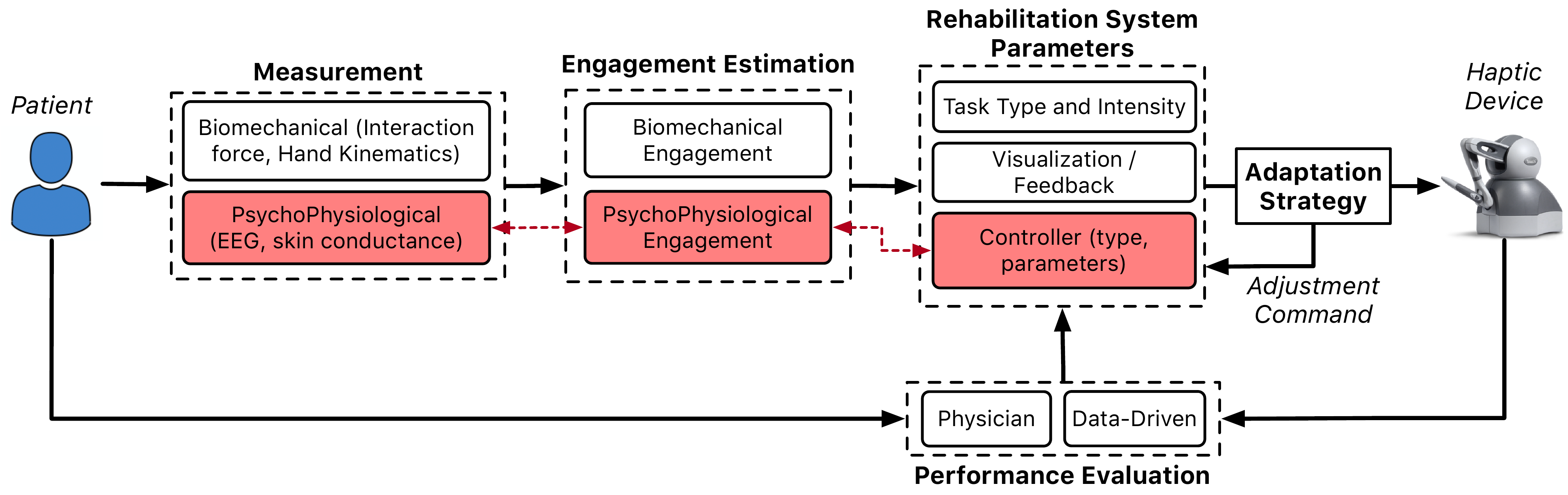}
    \caption{Typical framework of an adaptive haptic rehabilitation strategy. The highlighted boxes indicated the scope of this work which is studying the effect of haptic control strategy on mental engagement}
    \label{fig:framework}
\end{figure}

To maximize the patient's engagement during the rehabilitation, it is, therefore, necessary to adaptively modify the interaction parameters  \cite{simonetti2016multimodal}. Fig. \ref{fig:framework} demonstrates the general framework of such an adaptive rehabilitation system where interaction parameters (e.g., controller type/gains, visualization, feedback, task type, and intensity) are modified depending on the level of patient's engagement and performance to encourage his/her active involvement. 

To evaluate the rehabilitation performance as well as the engagement level, subjective (physician evaluations) and data-driven approaches can be used. It should be noted that the adaptation algorithm requires a proper understanding of the relationship between the subject's engagement and the system's interaction parameters. There are usually two types of engagement, Bio-mechanical and Psycho-physiological (cognitive), which are considered in rehabilitation systems for extracting the state of the patient.
Although the importance of both engagement types is very well understood for the success of rehabilitation \cite{simonetti2016multimodal, tsiakas2018taxonomy}, the majority of the adaptive rehabilitation techniques only consider the bio-mechanical measures such as applied force \cite{nordin2014assessment} or muscle activity \cite{wang2015emg} due to the ease of use. As a result, the relationship between bio-mechanical engagement and the rehabilitation control parameters are much better studied and understood \cite{simonetti2016multimodal} compared to cognitive engagement. This knowledge gap is mostly due to the difficulty in real-time estimation of mental engagement.

In this regard, electroencephalogram (EEG) activities have attracted much attention in quantifying the mental engagement as it provides a direct insight into the subject's cognition during rehabilitation \cite{berger2019current} with very good temporal resolution. Such quantification has been suggested as a continuous outcome during rehabilitation \cite{bartur2017eeg, li2016development} to change the interaction parameters to the optimum levels. In this context, to the best of our knowledge, there is still no study investigating the effect of control strategies type (EA or ER) on patient mental engagement during fine motor tasks. The scope of our work about the adaptive rehabilitation strategy is highlighted in Fig. \ref{fig:framework}. To reiterate, we are using EEG as a passive assessment method to quantify the engagement under different haptic control strategies, namely error augmentation and error reduction.

In this paper, we investigate the above-mentioned knowledge gap by studying the mental engagement of subjects during haptic-based fine motor writing tasks. For this purpose, our design of the experiment considers three types of haptic control strategies: ER, EA, and baseline (free control - no haptic assistance). For passive mental engagement quantification, we use the EEG-based index and explore two research questions i) which type of haptic control (EA v/s ER) evokes higher engagement in subjects, and ii) whether the subject's engagement depends on the hand (dominant or non-dominant) used in the haptic-based rehabilitation task. A Generalized Linear Mixed Model (GLMM) is used to conduct the statistical study on the effect of control strategy (EA v/s ER) and hand type (dominant v/s non-dominant) on the engagement level of the subjects. This study explores an objective assessment of the effect of haptic-control strategy on mental engagement during fine motor tasks. Understating the relation between mental engagement and haptic-control strategy can facilitate the design of adaptive haptic controllers that can promote the patient's engagement in home-based rehabilitation. Such controllers can leverage the mental engagement of the patients as a metric to adapt the level and type (EA v/s ER) of assistance being supplied to achieve better rehabilitation outcomes.

\subsection{Engagement in robotic rehabilitation}

In this section, we briefly highlight the importance of mental engagement in motor learning and the outcomes of robotic rehabilitation and then discuss the approaches with which the engagement can be evaluated.

There are multiple studies providing evidence that, in robot-assisted rehabilitation, passive movements are insufficient to improve motor recovery \cite{lynch2005} unless they are coupled by active movements and engagement \cite{hogan2006motions}.
For instance, Lynch et al. \cite{lynch2005} conducted randomized controlled trials with 32 patients who received continuous passive movements. They reported a positive trend in the motor learning of the patients but no significant differences when compared to patients performing therapist-supervised self-range of motion.

Moreover, it is suggested that motor rehabilitation is a form of learning \cite{WARRAICH2010, Dipietro2012} that can be enhanced by active patient participation/engagement \cite{WARRAICH2010}. 
Recent researches in robotic rehabilitation are therefore focused on maximizing the patient's engagement by providing assistance as needed and adapt the rehabilitation procedure according to the patient's intent \cite{marquez2016eeg, likitlersuang2018eeg, sullivan2017improving}. Blank et al. have provided a comprehensive review of the importance and the approaches for promoting patient engagement in robot-assisted stroke rehabilitation \cite{blank2014current}.

Gadi Bartur et al. \cite{bartur2017eeg} evaluated the relationship between the single-channel Brain Engagement Index (BEI) measured in terms of EEG and temporary functional changes induced during the standard rehabilitation sessions. The clinical study included 18 post-stroke patients with an average of 35 minutes treatment sessions followed by 30 second evaluation period. The study demonstrated that when BEI is higher, the temporary functional improvement due to the treatment session is also better. Trujillo et al. \cite{trujillo2017quantitative} assessed the relationship between qualitative EEG measures and the motor recovery outcome in chronic stroke patients in robot-assisted rehabilitation. Ten stroke patients with the upper limb deficit were recruited for the study. Clinical assessment was done by a physical therapist after one month of treatment using the Fugl-Meyer Assessment (FMA). The study showed that qualitative EEG measures were indeed correlated with motor recovery in chronic patients.

Along with the patient's mental engagement, the patient's intention to move is also used to assist or challenge as needed. For instance, Marquez-Chin et al. \cite{marquez2016eeg} used EEG-triggered functional electrical stimulation therapy to treat the upper limb, reaching the motion of a 64-year-old stroke patient. The study reported a clinically significant improvement in the Fugl-Meyer Assessment Upper Extremity as well as moderate improvement in Functional Independence Measure Self-Care subscore. On the same lines, Sullivan et al. \cite{sullivan2017improving} conducted a multi-year clinical study involving an EEG-based movement-intention detection for elbow flexion/extension rehabilitation. The results indicated that increasing the patient's engagement through intention detection enhanced the effectiveness of robotic rehabilitation system.

The above studies signify the importance of the patient's active mental engagement along with intensive repetitive movements (facilitated by a robot) on the outcome of robotic rehabilitation. Nonetheless, objective quantification of mental engagement is still not straight forward due to its' multi-dimensional nature.
Mental engagement includes different aspects of emotion, cognition, and motivation of the subject, making it an intricate and complicated feature and often associated with vigilance and alertness \cite{matthews2002fundamental}. To measure cognitive engagement, researchers have explored various modalities such as electrocardiography (ECG) \cite{Koenig2011}, galvanic skin response (GSR) \cite{berta2013electroencephalogram}, and electromyography (EMG) \cite{Zimmerli2013, pareek2019myotrackreal}, and pupillometry \cite{van2018pupil}. However, these modalities have certain drawbacks that limit their interpretability and scalability. For instance, the temporal resolution of ECG and GSR is poor and cannot relate directly to the stimuli. EMG signals cannot distinguish between passive and active movements from the patients \cite{goldberg2011predicting}. Eye features are sensitive to lighting conditions and are more relevant in visually oriented tasks \cite{mathot2018pupillometry}. Also, the pupillometry measures visual attention\cite{hartmann2014pupillometry} (the covert aspect of attention) that represent the level of visual information-gathering and provide vital information if the task difficulty is modulated by visual complexity. However, it contains less information about cognitive engagement when the task is not visual predominantly oriented.

Alternatively, recent advances in BCI have made it possible to measure cognitive engagement, which can potentially address the aforementioned problems \cite{esfahani2015adaptation}. In particular, using EEG has recently drawn significant attention \cite{Soekadar2015, bartur2017eeg, park2015assessment}. Many researchers have quantified mental engagement in terms of EEG-based engagement indices \cite{berka2007eeg, mcmahan2015evaluating}. These engagement indices are calculated using ratios of powers of different EEG frequency bands. The most common frequency bands used are Theta($\theta$: 4-7 Hz), Alpha($\alpha$: 8-13 Hz), and Beta($\beta$: 14-35 Hz) \cite{freeman1999evaluation}. 

Lubar et.al \cite{lubar1995evaluation} suggested the $\beta$/$\theta$ ratio averaged over all electrodes as an indicator of engagement. Pope et al. \cite{pope1995biocybernetic} and Freeman et al. \cite{freeman1999evaluation} improved that index by including the contribution of alpha rhythm as $\beta$/$(\theta + {\alpha})$ and suggested that it captured cognitive processes such as information gathering and attention. Berka et al. \cite{Berka2004} supported the use of $\beta$/$(\theta + \alpha)$ index by demonstrating that it correlates with sustained attention, information gathering, and visual scanning. Gevins and Smith  \cite{gevins1997high} introduced a new index $\theta/\alpha$, considering the theta and the alpha activities in the frontal and parietal cortices, respectively as a reflection of performance in demanding tasks. Yamada and Fumio \cite{yamada1998frontal} used only frontal mid-line $\theta$ power as an indicator of mental demand and focused attention. Within alpha band ($1/{\alpha}$), lower and higher alpha have been found to reflect attention and task processing, respectively \cite{klimesch1999eeg}. 
In our study, we used a Psycho-physiological baseline task to evaluate the robustness of the above-mentioned engagement indices. Out of five indices, $\beta/(\theta + \alpha)$ was selected for the subsequent analysis.

\section{MATERIALS AND METHODS} 
\label{sec:system}
\begin{figure}[!ht]
    \includegraphics[width= 0.5\linewidth]{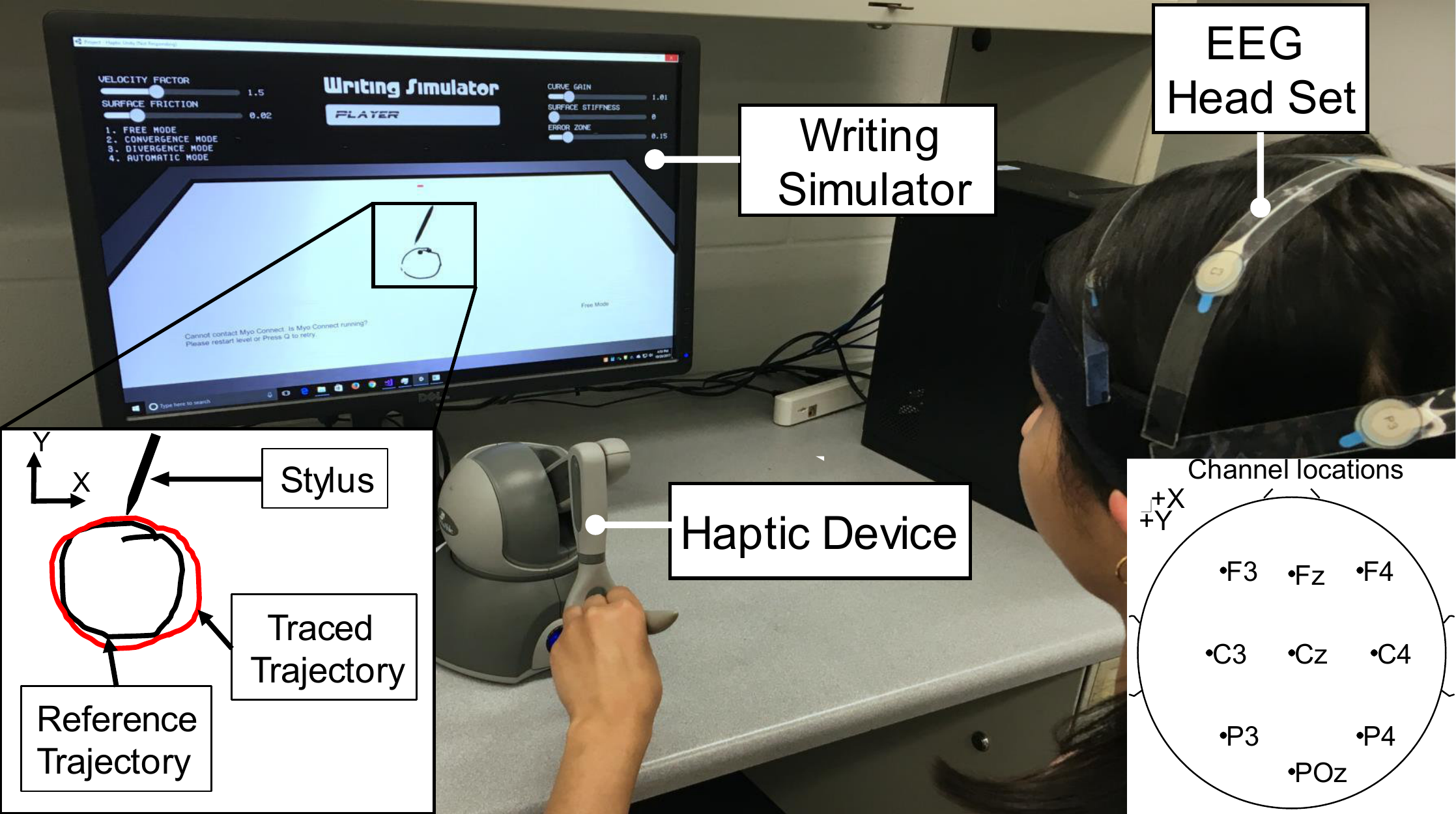}
    \caption{Experiment setup and writing simulator with haptics.}
    \label{fig:System}
\end{figure}

To regain writing skills, patients require kinesthetic assistance. In traditional rehabilitation programs, therapists provide such assistance to the patients, and as the patient's skills improve, the assistance is relaxed. In this paper, the same approach has been adopted for home-based robotic rehabilitation to recover a subject's writing skills. We have designed a writing simulation environment to study motor learning (trajectory tracking) of a subject with haptic assistance. As shown in  (Fig. \ref{fig:System}), the experimental setup consists of three main components: (1) a virtual environment to simulate the writing task, (2) a haptic device for assistance/resistance, and (3) an EEG headset to record the subject's cognitive activity.
\vspace{-20pt}
\subsection{Simulation Environment}
A writing simulation environment (Fig. \ref{fig:System}) is developed using the Unity3D interface, in which the end-effector of the haptic device acts as the writing stylus controller \cite{pareek2018development}. It allows the therapist to draw a template of interest (Fig. \ref{fig:System}) on the screen, which is then used as the reference trajectory to be followed by the patient.
In cases that the subjects deviate from the desired trajectory, a controlled force is applied by the haptic device to correct the trajectory. 
To achieve a high fidelity haptic rendering (thus better haptic assistance), the sampling rates of the simulation system  (100 Hz) and the haptic device (1 kHz) are synchronized by resampling the trajectory data using B-Spline interpolation. With this approach, a continuous function of the discretely sampled reference trajectory can be generated, which can be used for designing the controllers, as explained in the subsequent sections. Further, the B-Spline interpolation can be differentiated analytically to obtain a reference velocity. 

\begin{figure}[!h]
    \centering
          (a) \includegraphics[width=0.5\textwidth]{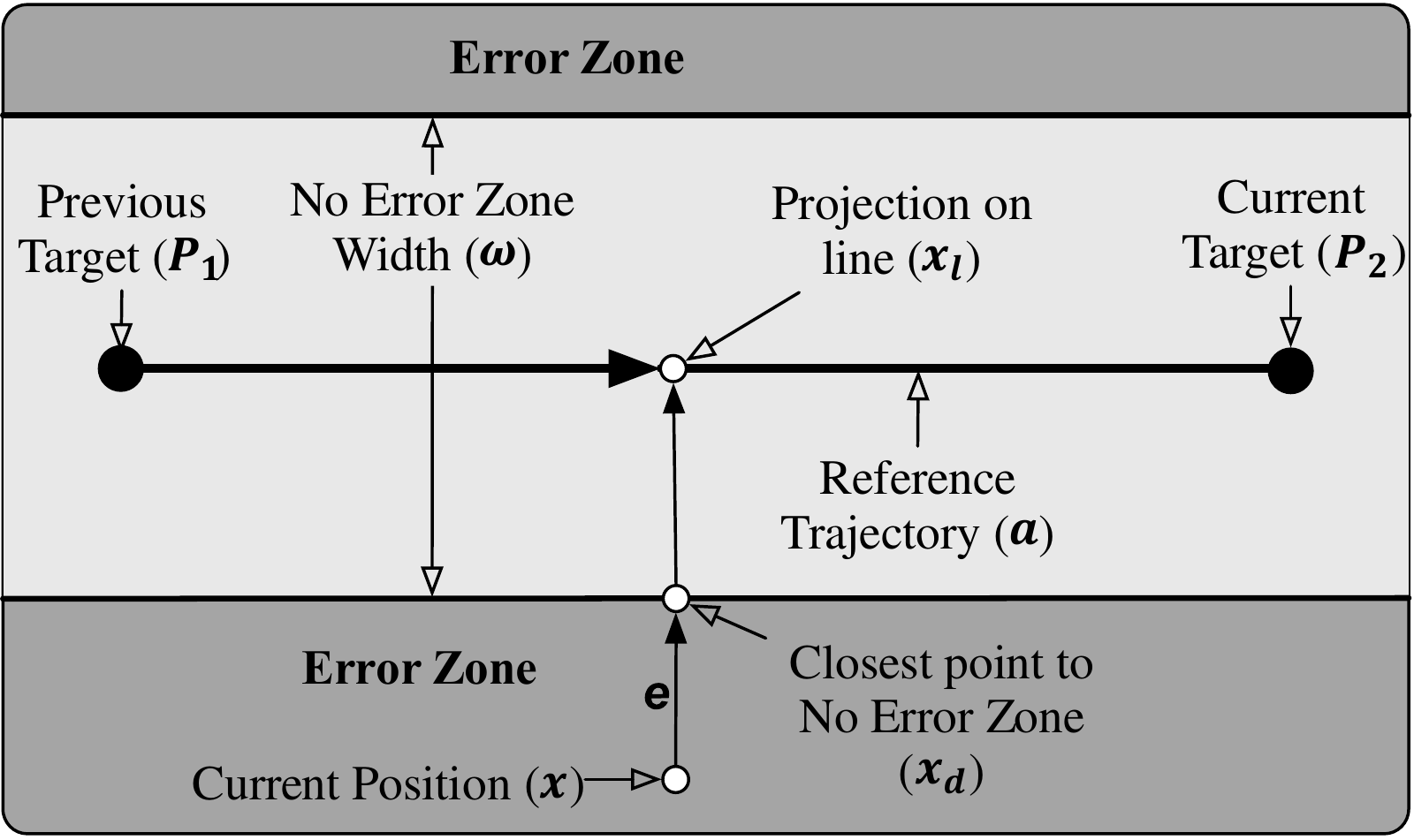}
      (b) \includegraphics[angle=90, width=0.15\textwidth, height = 2.25in]{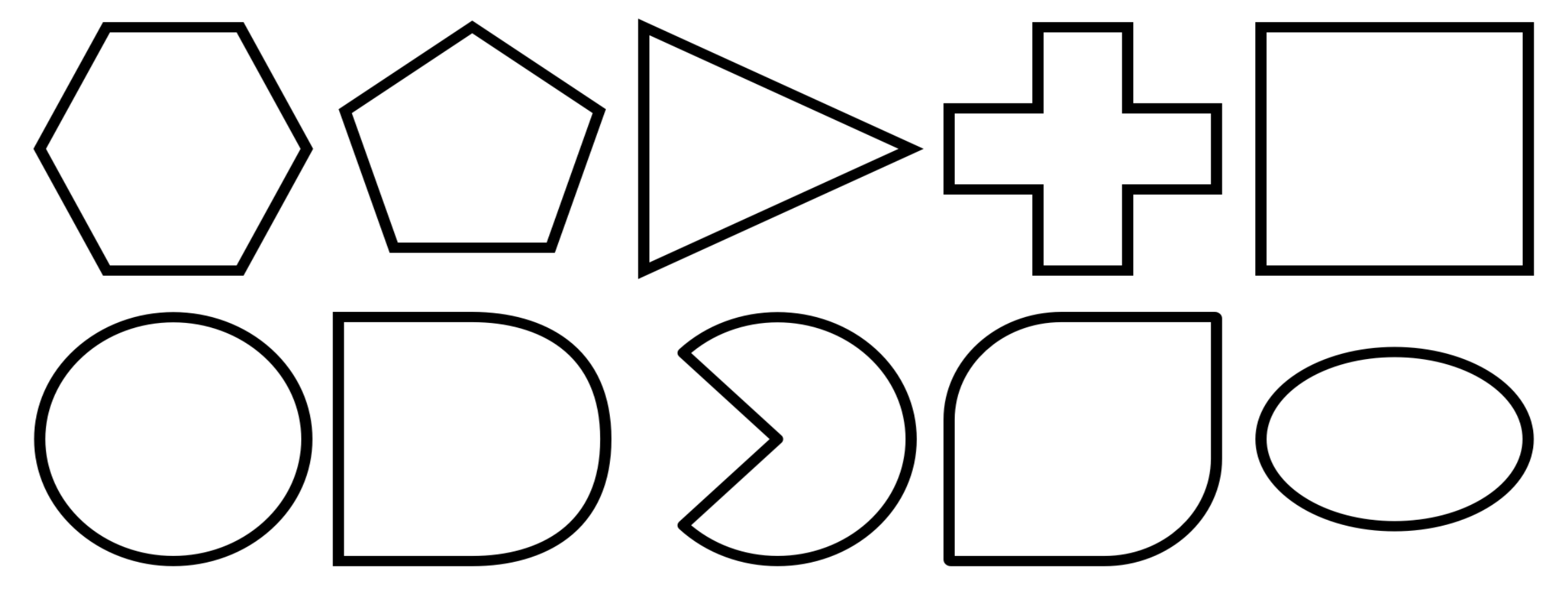}
    \caption{(a) Schematic representation of virtual no-error-zone and (b) templates used for the fine motor tasks.}
    \label{fig:errorZones} 
    \vspace{-5pt}
\end{figure}

To determine the tracking error ($\bf{e}$) between the desired position ($\bf{x_d}$) and the current position of the subject ($\bf{x}$), a virtual `no-error-zone' (Fig. \ref{fig:errorZones}a) is constructed around the trajectory. For instance, consider a simple case of tracking a straight line from one target ($P_1$) to another ($P_2$). The no-error-zone can be imagined as a rectangle with a width of $w$ and a length equal to the distance between the targets. While the end-effector is inside the rectangle, the tracking error is set to zero; otherwise, the error is defined as the minimum distance from the subject's position $(\bf{x})$ to the edge of the rectangle $(\bf{x_d})$. 

\subsection{Haptic Interface}
A 6 degrees-of-freedom (6 revolute joints - 3 actuated and 3 passive) Geomagic\textsuperscript{\textregistered} Touch\textsuperscript{TM} is used to provide force feedback to the user. The force feedback capability, a small form factor, lower cost, and 3-dimensional work-space make it a viable choice for home-based rehabilitation. The device is capable of applying a maximum force of 3.3 N and samples at a rate of 1000Hz. However, the range of force applied in limited. The simulation environment employs three control strategies: i) \textit{Free}, ii) \textit{Error-Reduction (ER) strategy}, and iii) \textit{Error-Augmentation (EA) strategy}. A detailed explanation of each control strategy is given in succeeding sections.
        
The haptic interaction with the simulation environment (haptic rendering) is modeled as a mass-spring-damper system. The robot provides an assistive or resistive force as the subject uses the system. The force command is generated based on a PD controller described by (\ref{eq:1}).
\begin{align}
    \textbf{e}(t)=\textbf{x}_d(t) - \textbf{x}(t)\notag\\
    \textbf{u}(t)=K_p \textbf{e}(t) - K_d\dot{\textbf{x}}(t)
    \label{eq:1} 
\end{align} 

Where $K_p$ and $K_d$ denote the proportional and derivative gains, respectively. $\textbf{u}$ is the control input provided by the robot to generate the haptic feedback. The control gains $K_p$ and $K_d$ determine the degree of robotic assistance or resistance which can be adjusted based on the subject's response. The derivative gain term ($K_d$) is set to a small positive value to simulate the sensation of moving through a lightly viscous environment. The desired velocity ($\mathbf{\dot{x}_d(t)}$) in (\ref{eq:1}) is set to zero, hence the second term is $K_d\dot{\mathbf{x}}(t)$.
\subsubsection{Free Strategy}
Free mode refers to the unassisted paradigm in the writing simulator. In this strategy, the robot applies no assistive/resistive force to the subject's hand. The subject is solely responsible for controlling the cursor motion. In this case, the proportional gain ($K_p$) of the control law is set to zero. This mode serves as a baseline for evaluating the subject's progress. The derivative gain ($K_d$) is fixed at 0.02 across all modes to simulate a friction sensation analogous to writing over a real piece of paper. 
\subsubsection{Error-Reduction (ER) Strategy}
In this strategy, as long as the subject remains inside the no-error-zone, the robot offers no assistance. If the subject moves outside the no-error-zone, the robot applies an assistive force towards the closest point, ${x}_d(t)$, on the rectangle's edge (Fig. \ref{fig:errorZones}a). The closest point on the spline is calculated as (\ref{eq:closest_point}). In certain cases, it may be possible that the closest point lies on an untraversed section of the trajectory. In such cases, the closest point is chosen by searching over the next $(k+n)^{th}$ points. $k$ is the index of the current closest point ${x}_d(t)$; and $n$ is the length of the search space (in this experiment $n=5$).  
\begin{equation}
    \textbf{x}_d(t) = \textbf{x}_l(t) - \frac{w}{2}(\textbf{x}_l(t)-\textbf{x}(t))
    \label{eq:closest_point}
\end{equation}
The control law according to error (${e}(t)$) and current position (${x}(t)$) to generate the assistive force for ER is given as (\ref{control}),
\begin{align}
    \textbf{u}(t)= 
    \begin{cases}
    K_p \textbf{e}(t) - K_d\dot{\textbf{x}}(t), & \text{if } d > \frac{w}{2} \\
    0, & \text{otherwise}  
    \end{cases}
    \label{control}
\end{align}

Where, $d$ is the Euclidean distance between the current point $\textbf{x}(t)$ and closest point on the line $\textbf{x}_l(t)$. The $K_p$ gain is set to a value of $1$. This value was chosen experimentally to ensure that sufficient assistance is provided to the subject without causing instability in the system.

\subsubsection{Error-Augmentation (EA) Strategy}
EA is similar in inception to ER with the exception that in the case of deviations from the error-zone, the haptic device forces the user away from the trajectory. In other words, $K_p$ is replaced by $-K_p$ in (\ref{control}). The subject then needs to guide the cursor back into the no-error-zone against the resistive force of the haptic device. The calculations for the closest point are the same as described for the ER. The $K_p$ gain is set to a value of $-1$. As with the previous case, this value was chosen experimentally to ensure that the resistance applied by the robot could be countered by the subjects without any discomfort and maintain controller stability.
\subsection{Human Subject Study}
In an experiment approved by the Institutional Review Board (IRB$\#$ 770128-1), ten subjects (six males and four females) were recruited from the students of University at Buffalo. Participants' age ranged from 19 to 26 years (Mean=24, SD=2.1), and they had normal or corrected to normal vision. 

\begin{figure}[!h]
    \centering
    \includegraphics[width=0.75\textwidth]{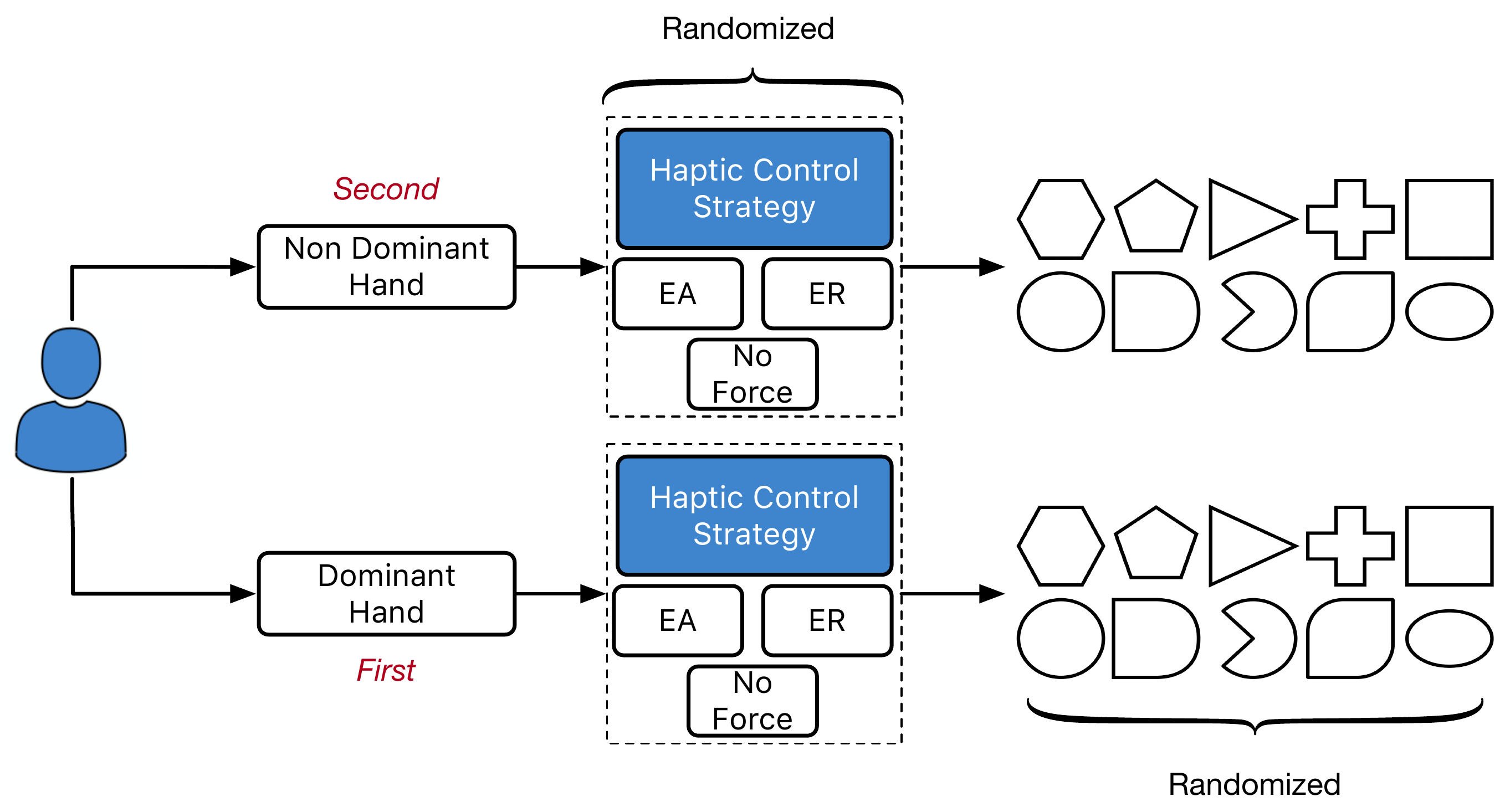}
    \caption{Experiment procedure flowchart. The order of the haptic control strategy and the choice of the shape were randomize for all subjects. All subjects were trained on their dominant hand and then the non-dominant one.}
    \label{fig:experiment_procedure} 
\end{figure}

The subjects were instructed to write one of the ten reference templates of the same size (Fig. \ref{fig:errorZones}b) using the simulator and traced it 10 times in each control mode. The templates were chosen and modified from the Visual-Motor Integration textbook \cite{Beery2005visual}. These simple templates were chosen to maintain the difficulty level of the task in an acceptable range and avoid any potential mental overload leading to the loss of engagement. All the subjects first performed the writing task using their dominant hand and then switched (with different figures) to their non-dominant hand. Under the dominant and non-dominant hand, all three control strategies were tested. The order of the control strategies and the templates were pseudo-randomized (Fig. \ref{fig:experiment_procedure}). 

Before the experiment, the subjects were briefed about the writing task. They practiced on the writing simulator using the Free strategy before the experiment. A reference trajectory was generated at the beginning of the experiment by the subject, which was used to calculate the errors with subsequent trajectories. During the experiment, the EEG activity of subjects was recorded. EEG signals were recorded using the B-Alert X10 wireless headset (Advanced Brain Monitoring$^{\textcopyright}$. Carlsberg, CA, USA) from 9 locations. These channels were F3, Fz, F4, C3, Cz, C4, P3, Poz, P4 according to the 10/20 international systems. We ensured the electrode impedance to be below 40 k$\Omega$.
\subsection{EEG Signal Analysis}
\begin{figure}[thb!]
    \centering
      \includegraphics[width=0.5\textwidth]{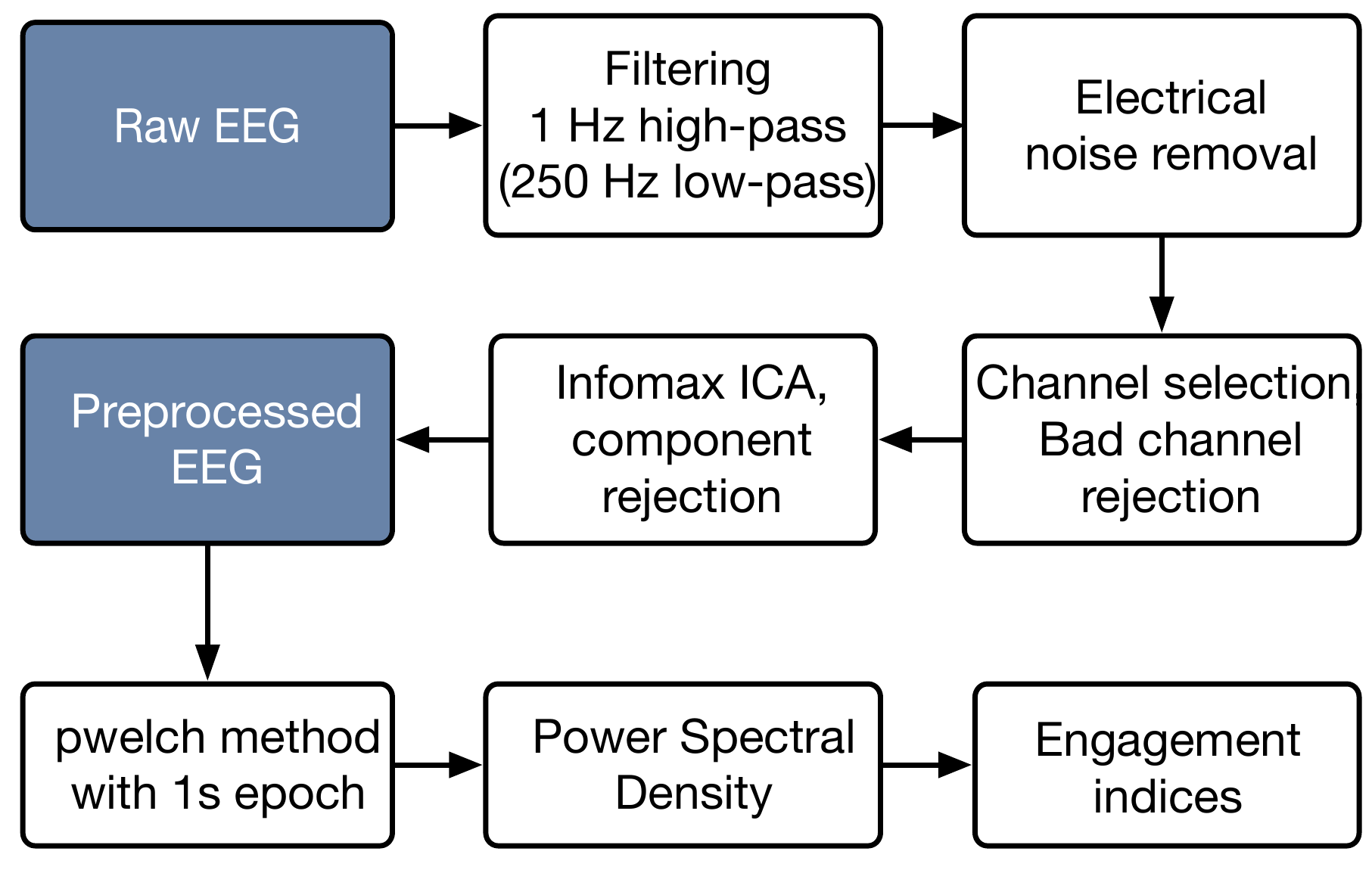}
    \caption{EEG analysis pipeline to calculate the engagement index.}
    \label{fig:Process}
\end{figure}

EEG signals were band-pass filtered (0.1-70 Hz) and then transmitted from the headset via a Bluetooth link to a nearby PC at a 256 Hz sampling rate. Artifacts caused by eye blinks and muscle contractions were removed using independent component analysis with the infomax algorithm in EEGLAB \cite{lee1999independent}. We visually examined 2-D scalp component maps to remove signal sources corresponding to eye movements and non-cognitive activities. After removal, the components were projected back to get an artifact-free EEG signal. 

Furthermore, relative and absolute power spectral densities were extracted using Welch’s method from 1-second epochs with a 50\% overlapping Hamming window. The features of 3 frequency bands, namely: theta (4-7 Hz), alpha (8-13 Hz), and beta (14-35 Hz) were used to extract five engagement indices commonly used in mental engagement analysis.
This workflow is presented in Fig. \ref{fig:Process}.

\subsection{Engagement Index Selection} 

\label{subsec:engagement-selection}
\begin{table}[h]
    \centering
    \tbl{The list of most common engagement indices used in the literature. The effectiveness of each index in distinguishing between the baseline experiments is shown in terms of p-values.}
    {
    \begin{tabular}{cll}
    \textbf{Index}                           & \textbf{Location}                                                 & \textbf{p-value} \\ \hline
    $\beta/(\alpha + \theta)$ & Avg. over all electrodes                                 & \textbf{$<$0.01}  \\ \hline
    $\theta/\alpha$         & Avg. frontal midline $\theta$ and avg. parietal $\alpha$ & \textbf{$<$0.01} \\ \hline
    $1/\alpha$            & Avg. over parietal electrodes                            & {0.025}  \\ \hline
    $\beta/\theta$      & Avg. over frontal electrodes                             & \textbf{$<$0.01}  \\ \hline
    $\beta/\alpha$          & Avg. over parietal and occipital electrodes              & 0.792 
        \end{tabular}
    \label{tbl:index_validation}
}
\end{table}
In this study, we considered five engagement indices (Table \ref{tbl:index_validation}), of which only one index was considered for subsequent analysis. Even though EEG-based engagement indices provide an objective measure of mental engagement, these measurements are dependent on individual differences and type of task. Hence, to select an engagement index that is robust towards the variation of a task and individual difference, a baseline task was conducted on twenty-two subjects (not participating in the primary human subject study), and the aforementioned five engagement indices were checked. The baseline included three tasks: Three-Choice Vigilance Task (3CVT), Eyes Open (EO), and Eyes Closed (EC) task corresponding to high engagement, low engagement, and relaxed wakefulness, respectively. For each index, we conducted a paired Student's t-test between the averaged scores of 3CVT and EO. The index with the significant discriminating power (lowest p-value) was selected as the engagement index for subsequent analysis. A significance level of $1\%$ (2-tailed) was used for all the comparisons. All the considered indices demonstrated statistically significant results $(p < 0.01)$ (refer Table \ref{tbl:index_validation}) except for $\beta/\alpha$. Among the indices, which showed significant results, ratio $\beta/(\alpha + \theta)$ was considered for the rest of the study as it is a widely used measure of engagement. 

\subsection{Statistical Analysis} 
To study the effect of haptic control strategy and hand type on the engagement level, we have used generalized linear mixed models (GLMM). It is an extension of the general linear model and considers both fixed and random effects. It is widely used for the analysis of grouped data, as it can model the differences between groups as a random effect.
GLMM formulation is given by equation \ref{eq:GLMM}, where the $\boldsymbol{\beta}$ are the fixed effect coefficients, and $\boldsymbol{u}$ are random effect coefficients and $\boldsymbol{X}$, $\boldsymbol{Z}$ are model matrices for fixed and random effects respectively.

\begin{equation}
    \label{eq:GLMM}
    \mathbf{ln(y)}=\boldsymbol{X} \boldsymbol{\beta}+\boldsymbol{Z} \boldsymbol{u}+\boldsymbol{\varepsilon}
\end{equation}

GLMM also allows the response variable, $\mathbf{y}$, to have different distribution rather than Gaussian. Throughout the study, we have used a natural log-transformed engagement index as the log-normal fitted the engagement index distribution well (see section \ref{subsec:engagement-statistics}). The log-transformation of $\mathbf{y}$ doesn't alter conclusions of the GLMM model because the natural log is a monotonic function.

In our study, we always consider each subject as a random effect on the intercept to account for individual differences. This consideration allows us to study the general trend in the main population from which the samples (subjects) are selected and not just the specific samples.

\section{RESULTS AND DISCUSSION}
Before presenting the results of the statistical analysis, it is crucial to determine if the subjects received two different levels of force distributions from the ER and EA controllers. This provides a validation of the haptic control implementation and also gives insights into the kinetic behavior of the dominant and the non-dominant hand. Moreover, we need to ensure that the levels of force chosen are such that the task is not too difficult or too easy for the subjects, which may lead to low engagement in either case. Also, we need to ensure that in our experimental design, we maintain the same level of bio-mechanical engagement (reflected in the interaction force and tracing speed and error) to avoid its confounding effect. Hence, in section 3.1, we first provide results of the kinesthetic aspects of our experiment: hand trajectory, force applied by the haptic device, error distribution in trajectory, and average speed of tracking to validate our design of the experiment. Section 3.2 concludes with GLMM results for the engagement index under the dominant and the non-dominant hand using different control strategies. 

\subsection{Kinesthetic Results of the Writing Task}
\begin{figure}[h]
    \centering
    (a)
     \includegraphics[width=0.72\textwidth]{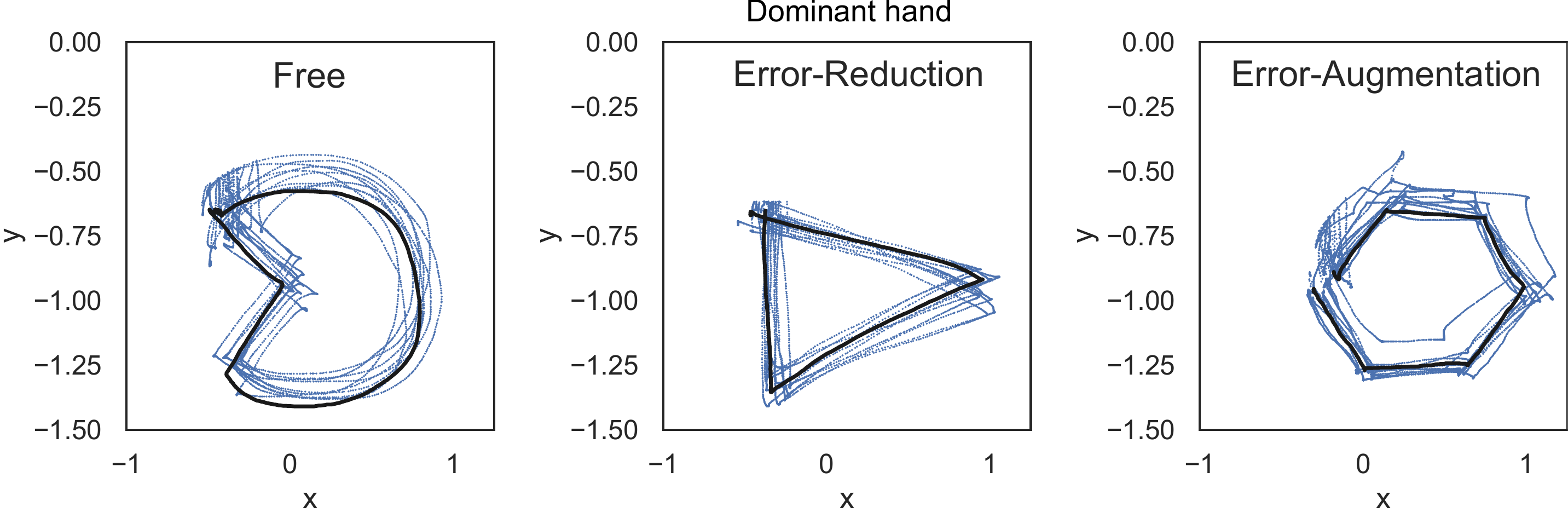}
     
    (b)
     \includegraphics[width=0.72\textwidth]{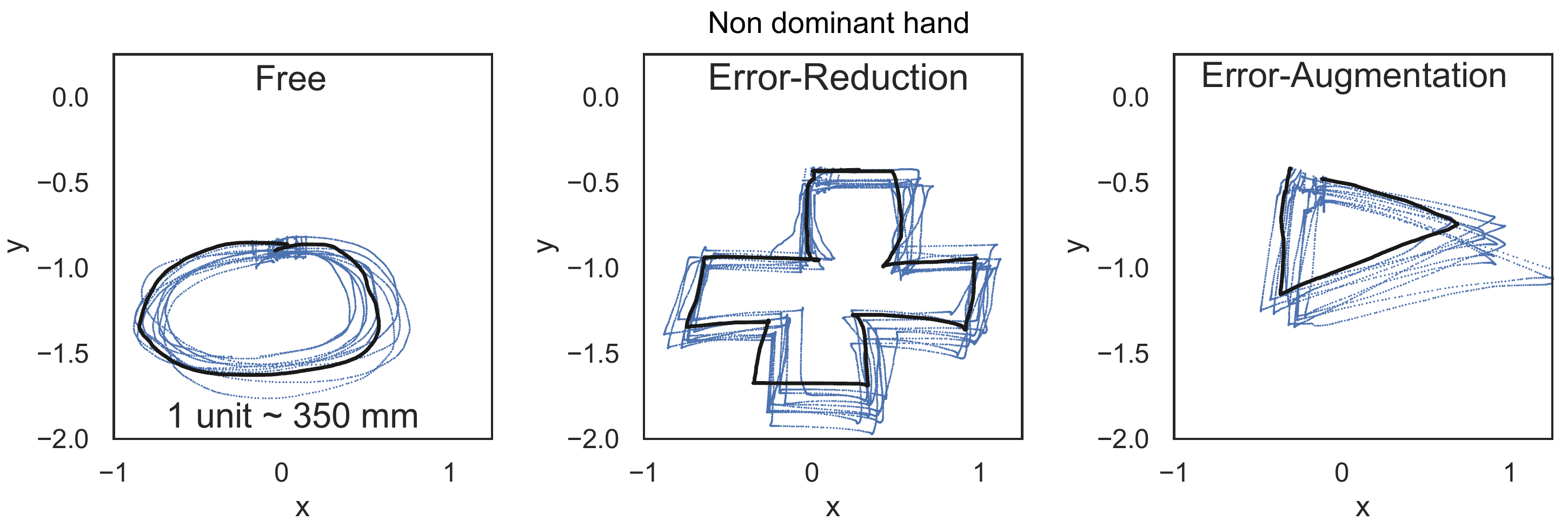}  
    \caption{Tracking data (1 unit $\approx$ 350 mm) for the (a) dominant and (b) non-dominant hand using different control strategies.}
    \label{fig:template_trace}
\end{figure}

Fig. \ref{fig:template_trace} shows the two-dimensional position data for one of the subjects performing the writing task under different haptic control strategies. The spread of traces is larger in the non-dominant hand when compared to the dominant hand. This is because the subjects are already familiar with fine motor movement in their dominant hands.

The EA strategy forces are distributed towards the positive side, and the ER strategy forces are distributed towards the negative side, which suggests that the control strategies behaved as intended (Fig. 7a). As the writing task is planar, we neglect the force in the z-direction.

The magnitude of EA and ER strategy forces are significantly different from the no-force condition under the dominant and non-dominant hand, which is expected. However, the magnitude of forces in EA and ER are not significantly different, which indicates that the main factor that is changing is the strategy type, not the magnitude of force itself. In terms of the average speed of tracking, there was no significant difference between different control strategies under dominant and non-dominant hands (Fig. 7b)). A plausible reason might due to simple drawing figures, and the force was applied only to correct the error in trajectory rather than guiding the subject along the trajectory.

\begin{figure}[h!]
    \centering
    \includegraphics{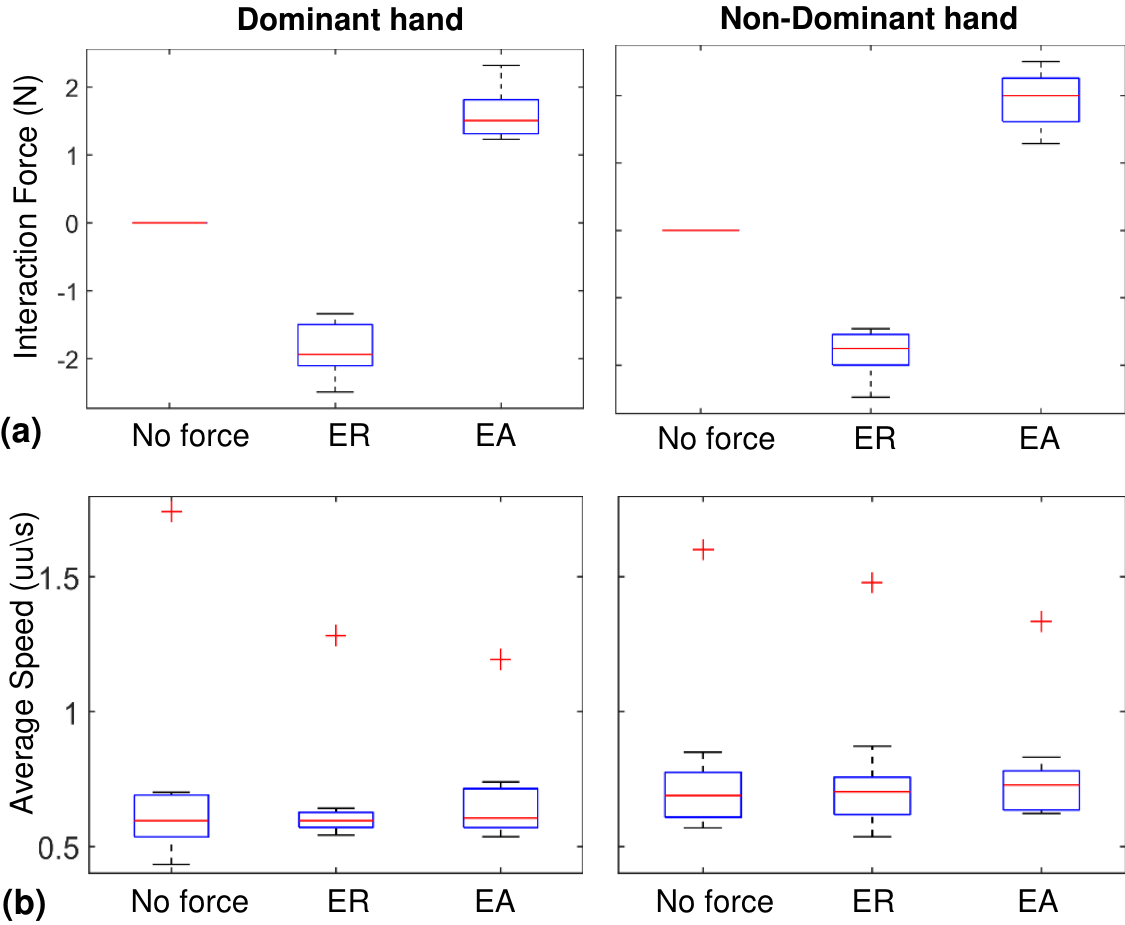}
    \add{\caption{\textbf{(a)} Comparison of the interaction forces using three control strategies for both dominant and non-dominant hand. All comparisons are significantly different. \textbf{(b)} Comparison of the average tracking speed using different control strategies under the dominant and non-dominant hand. All comparisons are statistically insignificant. The interaction force is measured in Newton (N), and the tracking speed is measured in Unity unit per second (uu/s).}}
    \label{hm3}
\end{figure}

\begin{figure}[htb!]
    \centering
      (a) \includegraphics[width=0.35\textwidth]{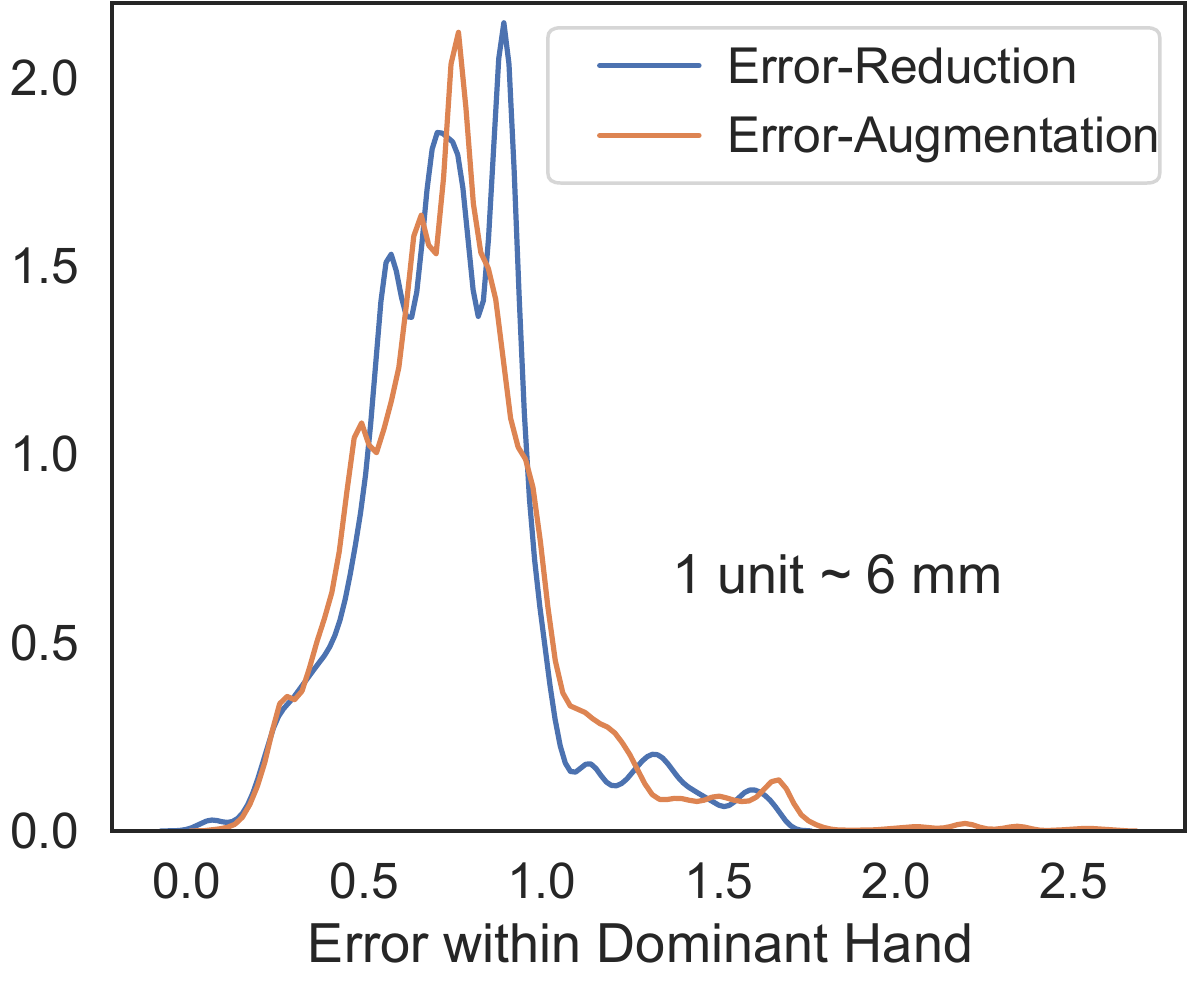}
      (b) \includegraphics[width=0.35\textwidth]{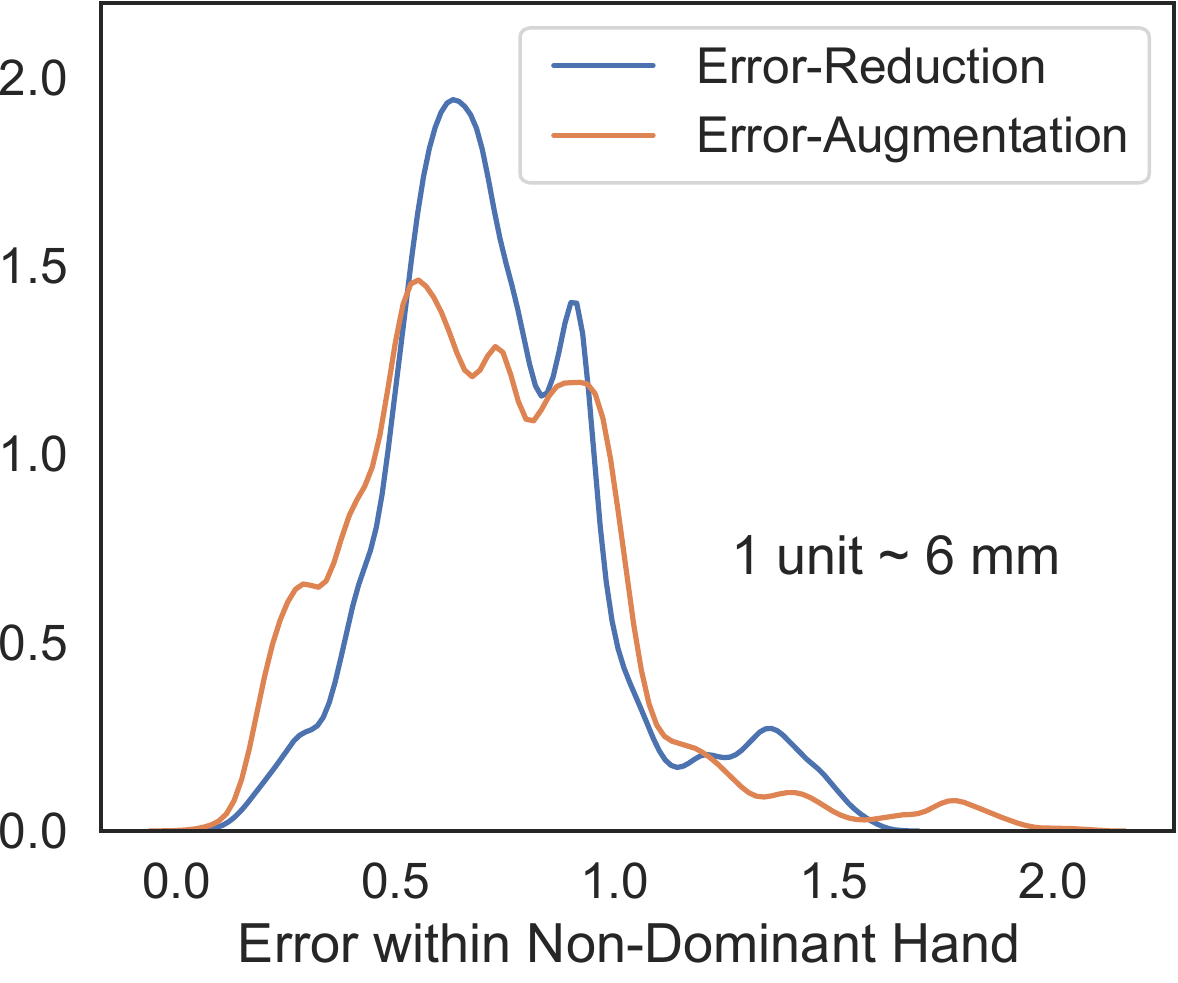}
    \caption{(a) Tracking error distribution for the dominant, and (b) Non-dominant hand under ER and EA control strategies.}
    \label{fig:error}
\end{figure}

In terms of tracking error, for the baseline task, the error is not significantly different from the tracking error in the dominant hand or non-dominant hand. A probable reason might be because healthy subjects participated in the experiment who are familiar with the writing task. Another probable reason might be the size of the shapes used for tracking. The shapes were not large nor complicated enough to be very challenging for healthy subjects to make a significant error. This constraint on the size of the shapes and the error zone comes from the design decision that we wanted to control the bio-mechanical engagement and study only active mental engagement. Also, there is no significant difference between the EA or ER within the dominant or non-dominant hand (Fig. \ref{fig:error}). The non-significant difference in tracking error under different control strategies also explains the non-significant difference in the magnitude of force distribution (Fig. 7a). 

In terms of variance (in tracking error), the variance in the non-dominant hand ($0.088$) is higher compared to the dominant hand ($0.06$). This may be due to the higher familiarity of the subjects with the fine motor control task using the dominant hand than the non-dominant hand. A similar error distribution across the two control strategies signifies that the control range chosen in the experiment is not too high to disengage or distract the subject. This is important because providing too much assistance can reduce the subject's involvement, and they may fail to learn the motor primitives needed to complete a task. Conversely, challenging the patient beyond a certain level might distract the subjects from the motor task itself, thereby leading to loss of engagement. 

\subsection{Statistical Analysis Results of Engagement Level}
\label{subsec:engagement-statistics}
For the general linear mixed models, it is important to consider the distribution of the response variable to fit the correct distribution. In this study, four different distributions, namely: Weibull, Gamma, Gaussian, and Log-normal, were fitted to the engagement index for each subject to identify the best distribution fit. To identify the goodness of fit, the Akaike information criterion (AIC) was used. AIC was calculated for each distribution fit and averaged over all the subjects (Table 2). Log-normal distribution on average (over subjects) gave the least AIC value, followed by Gamma and Gaussian (normal) distributions. Consequently, for the rest of the linear mixed models, Log-normal distribution was used. 

\begin{table}[h!]
 \tbl{Average (over subjects) Akaike information criterion (AIC) of different distributions fitted to engagement index.}{
  \centering
    \begin{tabular}{ccccc}
    \hline
    \textbf{Distribution} & \textbf{Weibull} & \textbf{Gamma} & \textbf{Gaussian} & \textbf{Log normal} \\
    \hline
    \textbf{Average AIC} & -1538.259 & -2463.218 & -1820.818 & \textbf{-2676.88} \\
    \hline
    \end{tabular}%
    }
 \label{table:aic_distribution}
\end{table}

Many researchers have shown that the beta, theta, and alpha rhythms are negatively correlated with the task engagement and alertness \cite{pope1995biocybernetic, freeman1999evaluation}. Concerning the engagement index, numerous studies have shown that beta oscillations have a vital role in attention-related processes \cite{kaminski2012beta}. The beta oscillations are normally associated with sensorimotor processing \cite{spitzer2017beyond}. The beta power (sensorimotor cortex) decreases during the preparation and execution of movement, but increases after the movement are complemented. Jenkinson and Perter \cite{jenkinson2011new} proposed that the level of beta activity is inversely proportional to the likelihood that new voluntary action will need to be processed and performed. They hypothesized that net dopamine levels and beta activity are inversely related \cite{knyazev2007motivation}. Theta oscillations in the engagement index $\beta/(\alpha + \theta)$, in general, are correlated with memory and emotional regulation (both positive and negative emotions). Also, the desynchronization of lower alpha activity is associated with the attention process \cite{knyazev2007motivation}. Jensen et al. \cite{jensen2002oscillations} suggested that frontal theta oscillations are observed during increased workload, indicating sustained attention to new information. Thus, an increase in $\theta$ activity indicates an increase in the attention process. Consequently, the index $\beta/(\alpha + \theta)$ is negatively correlated with the engagement level. The rest results from GLMM are interpreted based on this consideration. Moving forward, we provide a general model with all the factors. The factors considered are the control type, the magnitude of the force, and the hand type. The reason for considering only the magnitude of the force is that the sense of direction is already encoded in the control type. Within this model, the hand type is used as a random slope effect. This consideration accounts for dominant and non-dominant differences, as well as the variation of templates used in those hands. If any difference is found, a separate mixed model is constructed under the dominant and non-dominant hand to study the effect of control strategies on engagement. It should be noted that GLMM considers one level of each of the predictors as the reference. For the general model, all the comparisons are made taking dominant hand and no force control type as the references.

Under this general model, GLMM (Table 3) revealed a significant effect of controller type, hand type, and interaction of controller type with hand type under the non-dominant hand. Thus, we can conclude that depending on the hand type, control types affect the engagement level. Consequently, we constructed two different GLMM under dominant and non-dominant hands to study the effect of the control strategy.

\begin{table}[htbp]
  \centering
  \centering
   \tbl{GLMM results with all the factors and engagement index.}
   {
    \begin{tabular}{cp{12.335em}ccc}
    \hline
    \textbf{Effect} & \multicolumn{1}{c}{\textbf{Predictors}} & \textbf{Estimates} & \textbf{Confidence Interval} & \textbf{p-value} \\
    \hline
          & \multicolumn{1}{c}{Intercept} & -0.75 & [-0.96 ~ -0.59] & \textbf{$<$0.001} \\
    Main  & \multicolumn{1}{c}{ Error Augmentation (EA)} & 0.27  & [0.14 ~ 0.41] & \textbf{$<$0.001} \\
    Main  & \multicolumn{1}{c}{ Error Reduction (ER)} & 0     & [-0.16 ~ 0.16] & 0.987 \\
    Main  & \multicolumn{1}{c}{ Force} & 0.02  & [-0.13 ~ 0.18] & 0.749 \\
    Main  & \multicolumn{1}{c}{ Non Dominant} & 0.16  & [0.09 ~ 0.24] & \textbf{$<$0.001} \\
    Interaction &\multicolumn{1}{c}{ EA $\times$ Force} & -0.18 & [-0.39 ~ 0.01] & 0.065 \\
    Interaction &\multicolumn{1}{c}{ EA $\times$ Non Dominant} & -0.42 & [-0.63 ~ -0.21] & \textbf{$<$0.001} \\
    Interaction &\multicolumn{1}{c}{ER $\times$ Non Dominant} & -0.12 & [-0.37 ~ 0.11] & 0.281 \\
    Interaction &\multicolumn{1}{c}{ Force $\times$ Non Dominant} & 0.01  & [-0.21 ~ 0.24] & 0.886 \\
    \hline
    \multicolumn{2}{l}{Other interaction terms are not reported due insignificant effects.}
    \end{tabular}
    }
  \label{tab:generalmodel}
\end{table}

\begin{figure}[h!]
    \centering
      \includegraphics[width=\textwidth]{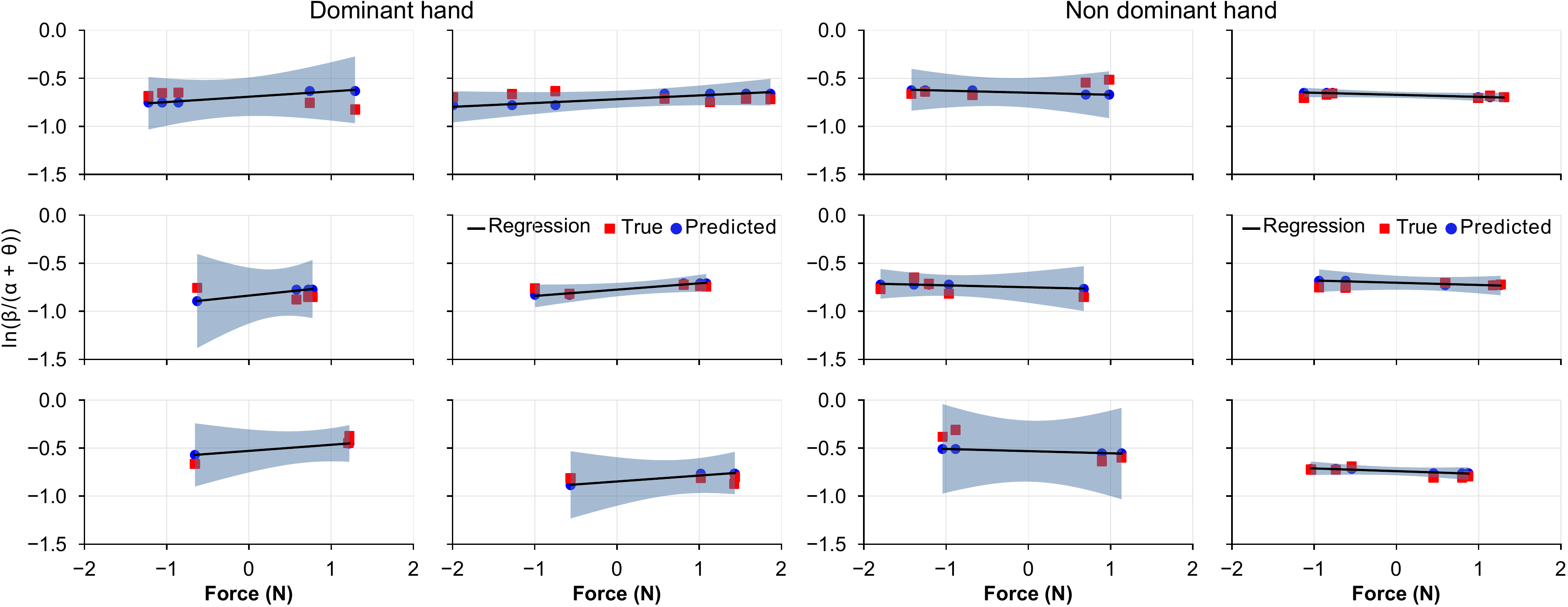}
    \caption{Trend of log-transformed engagement index with respect to the variation of force. The positive force indicates EA strategy and negative force indicates ER strategy.}
    \label{fig:mixed-prediction}
\end{figure}

Fig. \ref{fig:mixed-prediction} further highlights the above results under the dominant and non-dominant hand across different subjects. The regression shows a negative slope for the non-dominant hand and a positive slope for the dominant hand. This difference can be explained by the fact that the dominant hand is well versed in fine motor tasks such as writing due to extensive practice \cite{harris2006individuals}, resulting in less mental attention in executing simple motor tasks. In fact, using PET functional imaging studies, Grafton et al. \cite{Grafton2002} have shown the recruitment of widespread frontal and temporal regions in the brain during non-dominant hand motor learning which is identified as a source of higher consumption of attentional resources observed in ERP studies of non-dominant hand motor learning \cite{Rietschel2014}.

Moreover, the dominant hand is shown to have high motor units than non-dominant hand \cite{farina2003effect, tanaka1984comparison}. The non-dominant hand does not enjoy such an advantage; thus, it is expected to see some difference using different control strategies. This is highlighted further in the following GLMM model under the non-dominant hand.

The GLMM revealed a significant difference in engagement level in the non-dominant hand when the EA control strategy was used. The EA strategy decreased the engagement index by $0.14$ (Table \ref{table:engagement_index_regression}) units with respect to baseline (no force) strategy. Note that a decrease in engagement index signifies an increase in engagement level. With ER, the engagement index decreased by $0.11$. As discussed before, the $\beta/(\alpha + \theta)$ engagement index is negatively correlated with the engagement level; thus, EA strategy induces more engagement when compared to ER and baseline strategy under the non-dominant hand. We repeated the above model considering the ER strategy as a reference level; this reveals any differences between EA and ER strategy. There was a significant difference (p$<$0.05) under the EA strategy when compared to the ER strategy under the non-dominant hand.

\begin{table}[h]
    \centering
    \tbl{The results of linear mixed model of engagement index with respect to no-force control}
    {
    \begin{tabular}{ccccc}
    \hline
    \textbf{Hand Type}&\textbf{Predictors} & \textbf{Estimates} & \textbf{Confidence Interval} & \textbf{p-value}\\
    \hline
    &Intercept & -0.59 & [-0.71 ~ -0.47] & \textbf{$<$0.001}\\
    \textbf{Non Dominant}&Error Augmentation (EA) & -0.14 & [-0.18 ~ -0.11] & \textbf{$<$0.001}\\
    &Error Reduction (ER) & -0.11 & [-0.15 ~ -0.08] & \textbf{$<$0.001}\\
    \hline
    &Intercept & -0.77 & [-0.87 ~ -0.68] & $<$\textbf{0.001}\\
    \textbf{Dominant}&Error Augmentation (EA) & 0.14 & [0.11 ~ 0.18] & \textbf{$<$0.001}\\
    &Error Reduction (ER) & -0.01 & [-0.02 ~ 0.04] & 0.507\\
    \hline
    \end{tabular}%
    \label{table:engagement_index_regression}
    }
\end{table}

Under the dominant hand, using the EA, the engagement index increased by $0.14$ and decreased by $0.01$ under ER when compared to the baseline strategy. A line of reasoning for this trend can be that subjects are already familiar with fine motor tasks in their dominant hand. Hence, the chance of making a significant mistake is very low. So, any force applied might not lead to more engagement in the task.  Conversely, the non-dominant hand has less dexterity to perform fine motor tasks such as writing. Hence, a different level of engagement was observed only under the non-dominant hand.

\begin{figure}[h!]
    \centering
      (a) \includegraphics[width=0.45\textwidth]{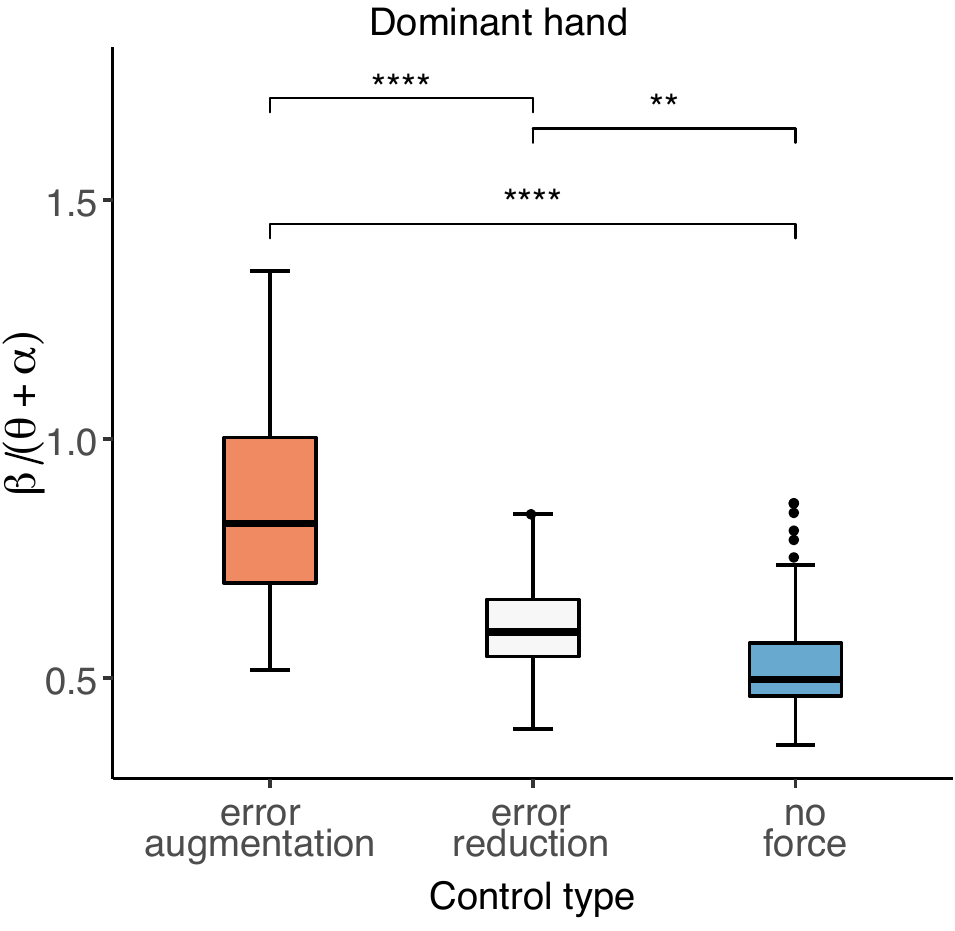}
      (b) \includegraphics[width=0.45\textwidth]{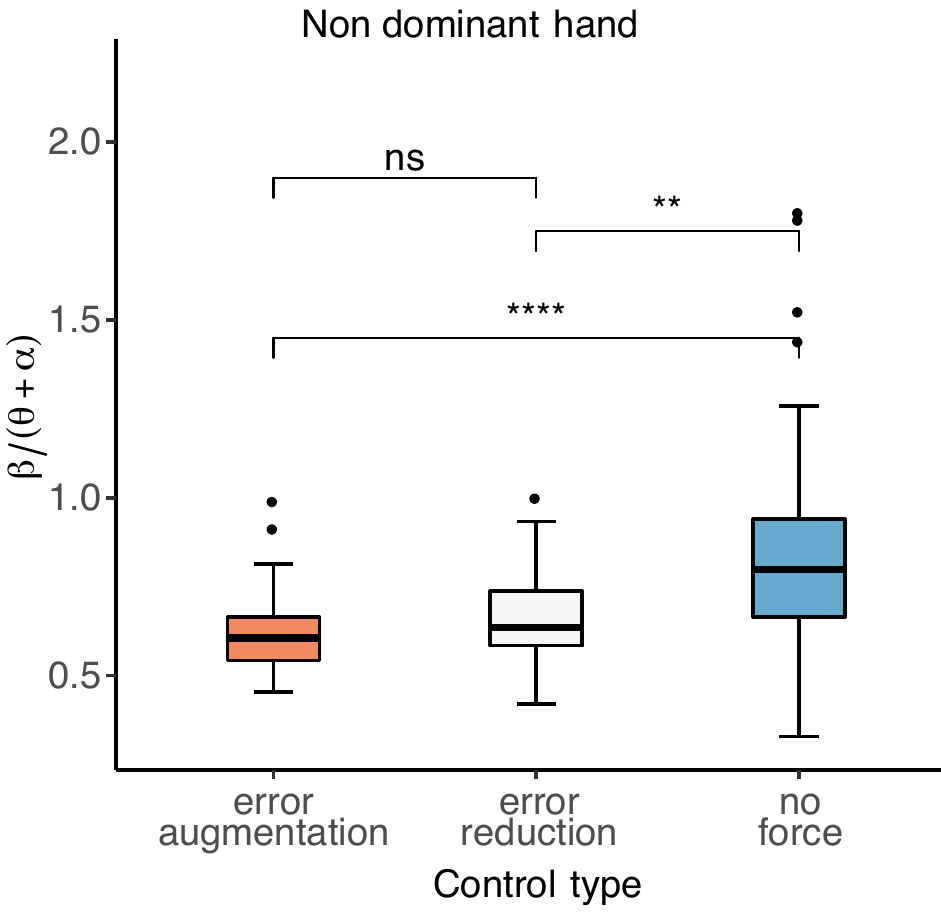}
    \caption{Significant differences between engagement indices using different control strategies under (a) dominant hand (b) non-dominant hand (The asterisk denotes significant difference, p $<$ 0.01).}
    \label{fig:significance}
\end{figure}

As a consolidation, Fig. \ref{fig:significance} shows the results for one of the subjects with different control strategies and the type of hand. Under the dominant hand, the engagement index in EA is significantly higher than the baseline (no force/free) strategy (Fig. \ref{fig:significance}a). However, in the non-dominant hand, the engagement index under EA and ER are significantly less than than the baseline strategy (Fig. \ref{fig:significance}b). Concretely,  EA induces more engagement when compared to the ER under the non-dominant hand. The observation of higher engagement in EA is also in agreement with the motor adaption principle, which indicates that neural signals that drive motor adaptation are generated by kinematic errors during movement \cite{marchal2014learning, albert2016neural, schmidt2018motor}. 

Another important measure of attention is Sensorimotor Rhythm (  SMR) in the sensorimotor region. The   SMR has been used in the study of psychomotor efﬁciency and attention-related tasks \cite {cheng2017higher}. In this context, John Gruzelier et al. \cite{gruzelier2006validating} provides a scoping review of neuro-feedback procedures using   SMR modulation and beta band inhibition for improving attention, mood, and memory. In this study, we used the average   SMR power (with log-transformed) calculated at the C3 and C4 between 13 Hz and 15 Hz to study the attention level under different haptic control strategies.

\begin{table}[htb!]
    \centering
    \tbl{The results of the general linear mixed model of  SMR power with respect to no-force control}
    {
    \centering
    \begin{tabular}{ccccc}
    \hline
    \textbf{Hand Type}&\textbf{Predictors} & \textbf{Estimates} & \textbf{Confidence Interval} & \textbf{p-value}\\
    \hline
    &Intercept & -27.54 & [-27.92 ~ -27.16] & \textbf{$<$0.001}\\
    \textbf{Non Dominant}&Error Augmentation (EA) & 0.11 & [0.05 ~ 0.18] &  \textbf{$<$0.001}\\
    &Error Reduction (ER) & 0.09 & [0.03 ~ 0.15] & \textbf{0.005}\\
    \hline
    &Intercept & -27.63 & [-28.04 ~ -27.21] & \textbf{$<$0.001}\\
    \textbf{Dominant}&Error Augmentation (EA) & 0.23 & [0.16 ~ 0.29] & \textbf{$<$0.001}\\
    &Error Reduction (ER) & 0.19 & [0.13 ~ 0.25] & \textbf{$<$0.001}\\
    \hline
    \end{tabular}
    \label{tabel:srm}
    }
\end{table}

\begin{figure}[ht!]
    \centering
      (a) \includegraphics[width=0.5\textwidth]{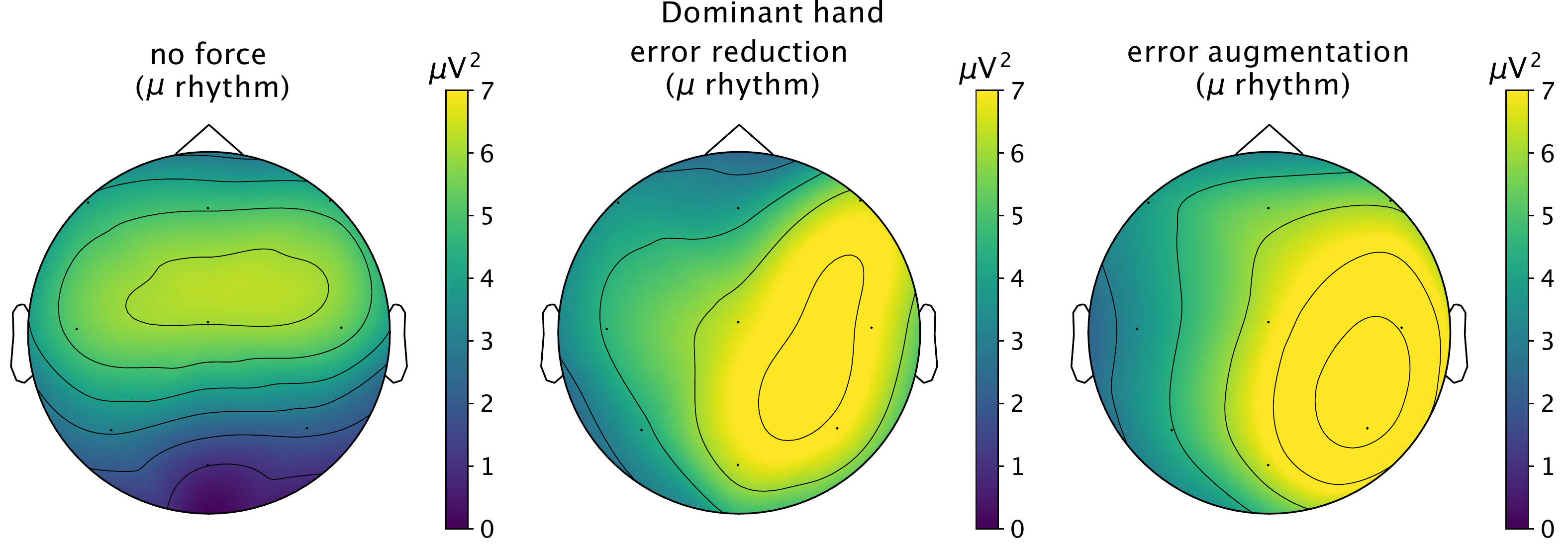}\\
      (b) \includegraphics[width=0.5\textwidth]{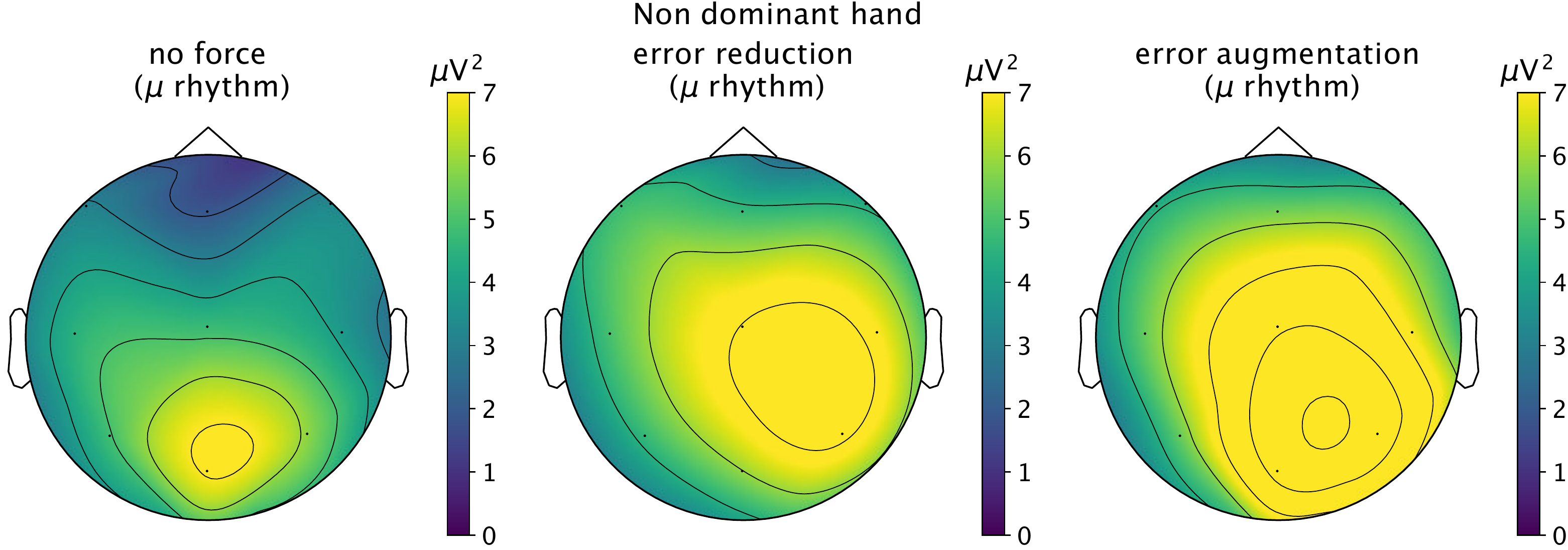}
    \caption{$\mu$ rhythm power over the scalp under different control strategies and hand type. The force significantly higher than the baseline condition under both dominant and non-dominant hand.}
    \label{fig:mu_rhythm}
\end{figure}

Table \ref{tabel:srm} presents the results of the general linear mixed model of sensorimotor rhythm power with respect to the baseline (no force). Under the non-dominant hand, the   SMR power using ER and EA control strategies is significantly more than the baseline (no force). However, under the EA strategy, the   SMR power slightly better. This results agrees with the engagement index results presented in Table \ref{table:engagement_index_regression} (Fig. \ref{fig:significance}). Under the dominant hand,   SMR of ER and EA strategies are significantly higher than the baseline (force). This result conflicts with previous results. Hence, a conclusive result cannot be drawn about the effect of control strategies under the dominant hand. The same can be seen in the $\mu$ rhythm topographic map for one of the subjects (Fig. \ref{fig:mu_rhythm}). The power is comparable between no-force condition under the dominant and non-dominant hand. However, the power is significantly more in error reduction as well as in error augmentation when compared to baseline, which reflects the results of the linear mixed model (Table \ref{tabel:srm}).

The limitations of the study are as follows. The controller gains were chosen based on the limitations of the haptic device. The haptic device used in the study cannot apply force more than 3.3 N. Consequently, a very large force could not be applied. As a result, we could not conclude any significant difference in bio-mechanical engagement parameters such as force and tracking error. Also, the participants considered in the experiments are healthy subjects who performed the experiments well, even with force from the haptic device. Consequently, the above-concluded results may differ if the system is used with clinical patients. Nonetheless, even though no significant difference in bio-mechanical engagement was not observed, a significant difference was observed in active mental engagement.

\section{CONCLUSION AND FUTURE DIRECTIONS}
This paper studies the effect of different control strategies on the mental engagement of subjects while performing fine motor control tasks using a haptic device with force feedback. An experiment is designed, where ten subjects perform a writing task using a haptic device that can either assist (error reduction) or challenge (error augmentation) the subject when they deviate from the reference trajectory. During the experiment, the subject's brain activity (EEG) is monitored to quantify mental engagement. The ratio of beta power (from all electrodes) and alpha + theta power is used ($\beta/(\alpha + \theta)$) as a measure of engagement.  With this setup, the study explores two research questions: i) which type of haptic control evokes higher engagement in subjects, and ii) whether the subject's engagement depends on the hand (dominant or non-dominant) used. General linear mixed models are used to identify the influence of different control strategies on the engagement level of the subject. When all the factors: hand type, control type, force magnitude was considered, hand type had a significant effect. Consequently, two GLMMs are constructed corresponding to the dominant and non-dominant hand. 

Under the non-dominant hand, the engagement of EA and ER is statistically higher than baseline control strategies. However, under the EA strategy, the engagement is slightly higher. Along with the EEG engagement index, Sensorimotor Rhythm (  SMR) power is also analyzed. The  SMR is significantly better for EA and ER strategy when compared to the baseline task under the non-dominant hand. The same trend is observed under the dominant hand also. However, the shapes that are considered in the study are comprised of simple arcs and lines, which can render the writing task too easy under the dominant hand. So, to arrive at more conclusive results under the dominant, more complex shapes need to be studied.

Even though General linear mixed models have enough power to signify the effect to control strategy, increasing the number of subjects can further bolster the study outcomes. Future work should investigate the relationship between higher levels of force feedback and engagement of the subject in the task. This is vital because making a task more challenging may cause the subjects to give up. Understanding such relationships facilitates adaptive assistance based on the mental engagement of the subject during fine motor tasks. 

\section{Acknowledgment}
This material is based upon work supported by the National Science Foundation under Grant No. 1502287 and 1502339.

\bibliographystyle{unsrt} 
\bibliography{references}
\end{document}